%% file: seke13p1-v13.tex
\documentclass[twocolumn]{article}
\usepackage{times}
\usepackage{epsfig}
\usepackage{vmargin}
\usepackage{latex8}
\usepackage{amssymb}
\usepackage{amsmath}
\usepackage{breqn}
\usepackage{url}
\usepackage{subfig}
\setpapersize{USletter}
\setmarginsrb{2.0cm}{2.0cm}{2.0cm}{1.5cm}{0pt}{0mm}{0pt}{8mm}

\newcommand{\sqm}{\begin{pmatrix}}

\begin{document}
\title{Understanding Common Perceptions from Online Social Media}

\author{
Derek Doran and Swapna S. Gokhale \\
{Dept. of Computer Science \& Engineering}\\
{University of Connecticut, Storrs, CT, 06269}\\
{\{derek.doran,ssg\}@engr.uconn.edu}
\and
Aldo Dagnino \\
Industrial Software Systems\\
ABB Corporate Research, Raleigh, NC, 27606\\
aldo.dagnino@us.abb.com
}
\maketitle

\begin{abstract}
Modern society habitually uses online social media services 
to publicly share observations, thoughts, opinions, and beliefs at any time and 
from any location. These geo-tagged social media posts may provide aggregate 
insights into people's perceptions on a broad range of topics across 
a given geographical area beyond what is currently possible through services 
such as Yelp and Foursquare. This paper develops probabilistic 
language models to investigate whether collective, 
topic-based perceptions within a geographical area can be extracted from 
the content of geo-tagged Twitter posts. The capability of the methodology is 
illustrated using tweets from three areas of different sizes. 
An application of the approach to support power grid restoration
following a storm is presented.
\end{abstract}

\section{Introduction and Motivation} 
\label{sec:intro}
Online social media services are now deeply rooted in our modern culture 
and people routinely turn to these services to share their thoughts
and opinions. These frequent updates 
can provide tremendous insights into how people perceive the 
world around them. A significant portion of these updates
are shared via smartphones and mobile devices, and 
hence, have location information embedded in them. 
This geo-tagging offers a unique 
opportunity to understand how the content or {\em what} of the posts
is influenced by the location or from {\em where} the posts are 
shared~\cite{brandom08}. Such linking of ``what'' to ``where'' 
can be used to support many geographic information retrieval 
systems~\cite{andogah10}. Commercial agencies can also use this 
association between perception and location to tailor their marketing 
strategies to geographic demands~\cite{lesser86}. 

Presently, geo-tagged social media posts are linked to
specific businesses using services such as 
Yelp\footnote{\url{http://www.yelp.com}} and 
Foursquare\footnote{\url{http://www.foursquare.com}}. 
Through this linking, people can share their reviews and experiences 
and alert friends to their whereabouts. Opinions about specific 
businesses, however, may not offer insights into how people 
feel about the underlying abstract notions or topics. 
For example, reviews about specific fast food restaurants cannot
indicate whether people in the area like to eat fast food, or 
that eating fast food is popular. Instead, posts that talk 
about fast food generally, with or without reference to specific restaurants, 
provide clues about area-wide perceptions on the topic of 
fast food. 

This paper proposes a methodology that uses social media 
posts to identify localities where a specific topic-based perception runs 
strong. Partitioning a geographic area into non-overlapping
sub-areas, the methodology trains probabilistic 
language models over posts from these sub-areas. This ensemble of models 
is then queried with a phrase defining a topic-based perception to
identify sub-areas where that perception runs strong. 
Illustrations using Twitter feeds from three areas of 
vastly different sizes, population densities, and other characteristics
show that despite the diversity, the methodology can identify sub-areas 
with strong perceptions for several common topics. The 
paper concludes with an industrial application of the methodology 
to support efficient power recovery following a major weather storm. 

The paper is organized as follows: Section~\ref{sec:model}
presents our methodology. Section~\ref{sec:data}
describes Twitter data. Section~\ref{sec:ill}
illustrates the methodology. Section~\ref{sec:appl} applies it
to storm damage response. Section~\ref{sec:rr} 
compares related work. 
Conclusions and future directions are in Section~\ref{sec:conc}.

\section{Locating Perceptions}
\label{sec:model}
In this section, we motivate and present the methodology 
for finding perceptions.

\subsection{Defining Perceptions}
People's perceptions about various topics may be embodied in their 
spoken language and now in their social media posts. 
These topic-based perceptions may be influenced by 
the general characteristics of an area and also by 
specific local features. For example, although people across 
New York City may frequently talk and
post about traffic congestion and delays, this issue is unlikely to be 
on their minds as they stroll through Central Park. 
Because it is impossible to exhaustively define all topics and their 
perceptions, we propose a flexible approach which models the language of 
the social media posts for each sub-area within a given area. These 
language models can then be queried with multi-word phrases which 
define a perception about a given topic to identify sub-areas which 
strongly represent that topic-based perception. For example, to identify 
perceptions of stressful (slow) traffic, we can use the query ''hate traffic''
(''traffic is slow''). Note that sub-areas with stressful traffic may or 
may not overlap those with slow traffic. 

\begin{figure*}
\vspace*{-0.05in}
  \centering
  \subfloat[NYC: Local-level]{\label{fig:nyc}\includegraphics[scale=0.28]{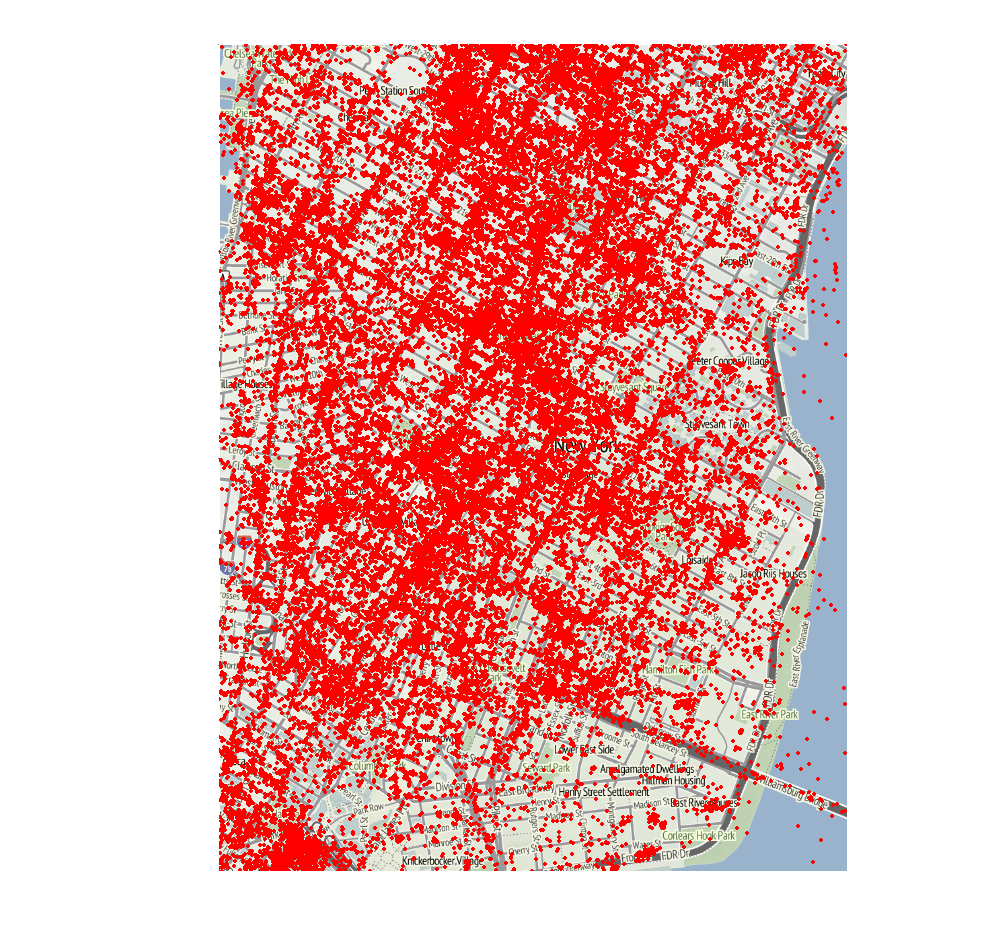}}
  \subfloat[DC: District-level]{\label{fig:dc}\includegraphics[scale=0.28]{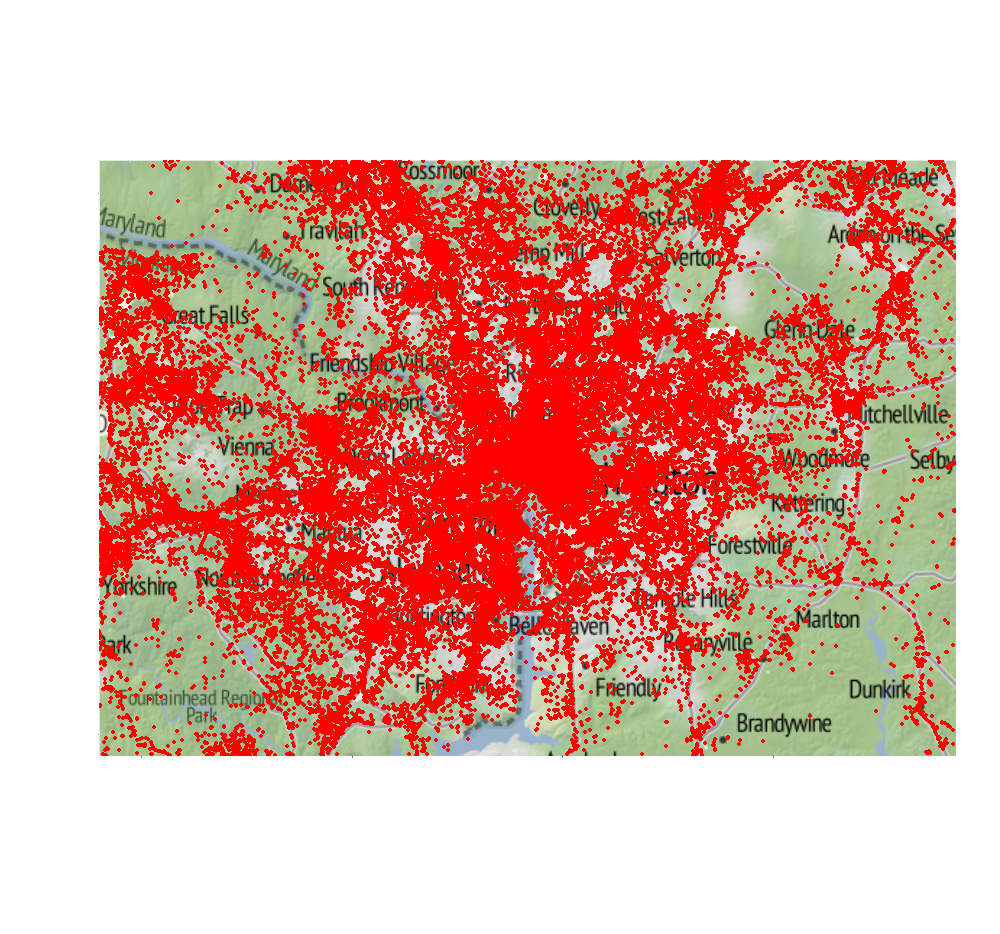}}
  \subfloat[CT: Region-level]{\label{fig:ct}\includegraphics[scale=0.28]{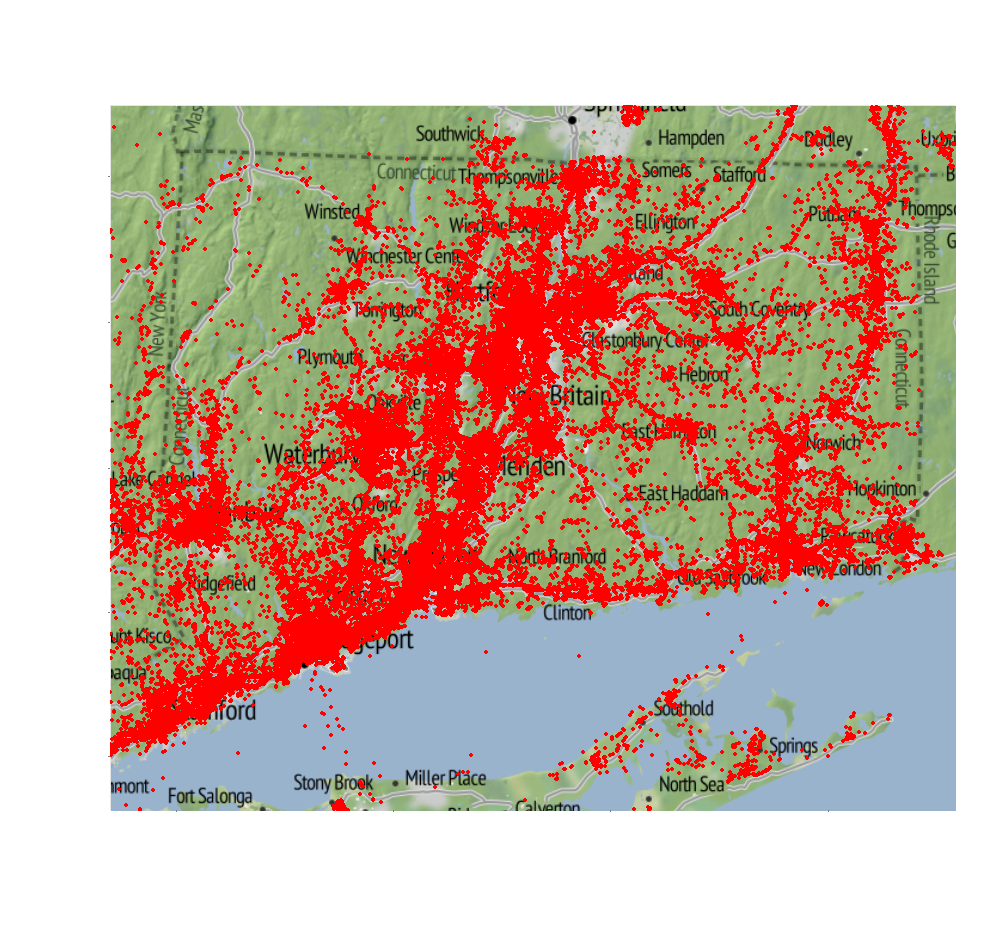}}
  \caption{Tweet Distribution in the Areas}
  \label{fig:locs}
\vspace*{-0.05in}
\end{figure*}

\subsection{Specifying Language Models}
A language model captures the features of 
the written language in a collection of documents or
a training corpus. It defines a probability distribution over 
all $n$-grams, where an $n$-gram is an ordered sequence of $n$ 
words $(w_1,....,w_n)$. We define language models for 
non-overlapping sub-areas $\ell_i$ which comprise an 
area $\mathcal{L}$. The maximum likelihood estimate of 
an $n$-gram, computed over a corpus of social media posts
within $\ell_i \in \mathcal{L}$, is given by~\cite{bahl83}: 
$$P_{\ell_i}(w_1,....w_n) = \frac{c(w_1,...w_n)}{c(w_1,...,w_{n-1})}$$
where $c(.)$ is the number of times the sequence appears in the posts. 
The probability that the language within a sub-area 
generates a phrase $T = (w_1,...,w_k)$ is computed as the 
product of the probabilities of the $n$-grams that comprise $T$:
\begin{equation}
P(T |\ell_i) = \prod_{j=n}^k P_\ell(w_{j-n+1},...,w_{j}) \nonumber
\label{eq:one}
\end{equation}

Contextual information increases with $n$ as higher order 
sequences of words are considered. However, because the frequency 
that larger sequences appear in social media posts is very low, 
prevalent language models over these posts restrict to the lowest order 
unigrams ($1$-grams), which model distinct words independent 
of their order~\cite{wing11,chandra11,li11}. Unigrams  
cannot model perceptions because these 
must be understood in the context of some topic or subject. 
For example, unigrams trained over ``I love driving'' 
and ``I hate driving'' will model the perceptions of ``love'' 
and ``hate'' but will not associate them with 
the topic of ``driving''. Bigrams, on the other hand, 
can model the perceptions ``I love'', ``love driving'', 
``I hate'', and ``hate driving" and associate them 
with a person (``I'') and the act of driving. Thus, 
unigrams can only recognize ``love'' or ``hate'', while bigrams actually 
identify {\em what} is being loved or hated. However, because 
social media posts may contain numerous distinct bigrams 
presenting unique thoughts on various topics, the estimates of bigrams  
over these posts may be inaccurate. To improve
this accuracy, we model the probability of bigram estimates using a 
linear interpolation of both the bigram and unigram estimates. 
This interpolation compensates for the low count 
of a bigram (e.g. ``love driving'') by incorporating the 
expected higher count of the unigram ``driving''. Thus, for a sub-area 
$\ell_i$, the probability of observing the bigram $(w_{j-1},w_{j})$ is 
given as: $$P_{\ell_i}(w_{j-1},w_j) = \lambda_1 c(w_{j-1},w_j) / c(w_{j-1})
+ \lambda_2 c(w_j) / |W(\ell_i)|$$ where $\lambda_1 + \lambda_2 = 1$,
$|W(\ell_i)|$ is the number of distinct words in all posts in
$\ell_i$ and $c(w_j) / |W(\ell_i)|$ is the estimate of the unigram
that completes the bigram~\cite{bahl83}. 

We use {\em smoothing} to compensate for the probability 
of future unseen bigrams, which allocates some of the probability of the 
training bigrams to those that are as yet unobserved~\cite{chen96}. The 
Modified Kneser-Ney (MKN) smoothing algorithm~\cite{kneser95}
is chosen because of its superior performance with interpolated language 
models~\cite{chen96}. The MKN algorithm subtracts a constant $\hat{d}$ from 
the observed frequency of every known bigram. It then estimates 
the likelihood that an unknown bigram $(w_{j-1},w_j)$ will appear using a 
modified estimate of the unigram $w_j$, where only the number 
of {\em distinct bigrams} that $w_j$ completes is considered:
$$P_c(w_j) = \frac{ |\{w: c(w,w_j) > 0\}|}{\sum_{v}|\{w: c(w,v) > 0\}|}$$
and then weighing this proportion by the probability 
mass $\lambda(w_{j-1})$ taken from the known bigrams: 
$$\lambda(w_{j-1}) = \frac{\hat{d}|\{w : c(w_{j-1},w) > 0\}|}{c(w_{j-1})}$$
Thus, under MKN smoothing the probability of observing a bigram becomes:
\begin{equation}
  P_\ell(w_{j-1}, w_j) = \frac{\max(c(w_{j-1},w_j)-\hat{d},0)}{c(w_{j-1})}
    + \lambda(w_{j-1})P_c(w_j) \nonumber
\label{eq:two}
\end{equation}
If $(w_{j-1}, w_j)$ is unknown, the probability is just given 
by $\lambda(w_{j-1})P_c(w_j)$, and if it is known, the probability is
given as a linear interpolation of the modified bigram and unigram estimates. 
Note that the modified unigram estimate $P_c(w_j)$ is superior to 
$c(w_j) / |W(\ell)|$ because 
under $P_c(w_j)$ words that appear frequently but within few
distinct contexts will not strongly influence the probability of
the bigram. We estimate $\hat{d}$ such that the log-likelihood 
that the model generates a given bigram is maximized:
$$ \hat{d} = \arg\max_d \sum_v c(v,w_j) \log P_\ell(v,w_j)$$
This has a closed form approximation depending on whether 
$c(w_{i-1},w_i)$ is equal to $1$, $2$, or $\geq 3$~\cite{sundermeyer11}. 
Using these approximations, we set 
$\hat{d}$ equal to $d_1, d_2$, or $d_3$ respectively:
\begin{eqnarray}
d_1 = 1-\frac{2n_2n_1}{n_1(n_1 + 2n_2)} \nonumber \\ 
d_2 =  2-\frac{3n_3n_1}{n_2(n_1 + 2n_2)} \nonumber \\
d_3 =  3-\frac{4n_4n_1}{n_3(n_1 + 2n_2)} \nonumber
\end{eqnarray}
with $n_i$ is the number of bigrams with frequency $i$.

The ensemble of language models, one for each sub-area, is 
then queried to compute the probability that a 
phrase $T$ is generated from a sub-area $\ell_i$ using Bayes rule:
$$
P(\ell_i | T) = \frac{P(T | \ell_i)P(\ell_i)}{\sum_j P(T | \ell_j)P(\ell_j)}
$$
$P(\ell_i)$ is the prior probability that a social media post is 
from sub-area $\ell_i$ and is given by $N(\ell_i) / N(\mathcal{L})$.
$N(\ell_i)$ is the number of posts in $\ell_i$ and $N(\mathcal{L})$ is 
the total number of posts in the entire area $\mathcal{L}$. Finally, we 
we define define $P(T | \ell_i)$ as: 
$$ P(T |\ell_i) = \prod_{j=2}^k P_{\ell_i}(w_{j-1}, w_j)$$

\input{seke13p2-v12}

%% file: seke13p2-v12.tex
\section{Data Description}
\label{sec:data}
\begin{figure*}[!ht]
  \centering
  \subfloat[``Restaurant"]{\label{fig:rest1}\includegraphics[scale=0.28]{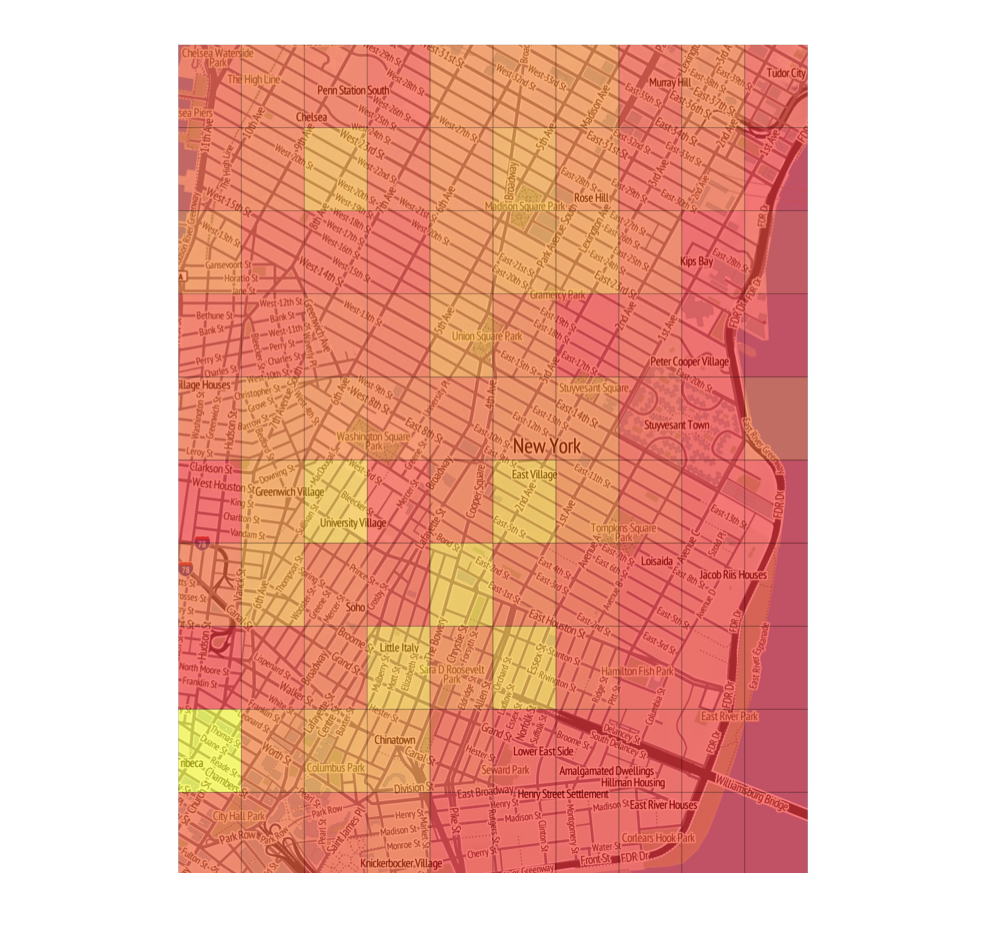}}
  \subfloat[``Italian restaurant"]{\label{fig:rest2}\includegraphics[scale=0.28]{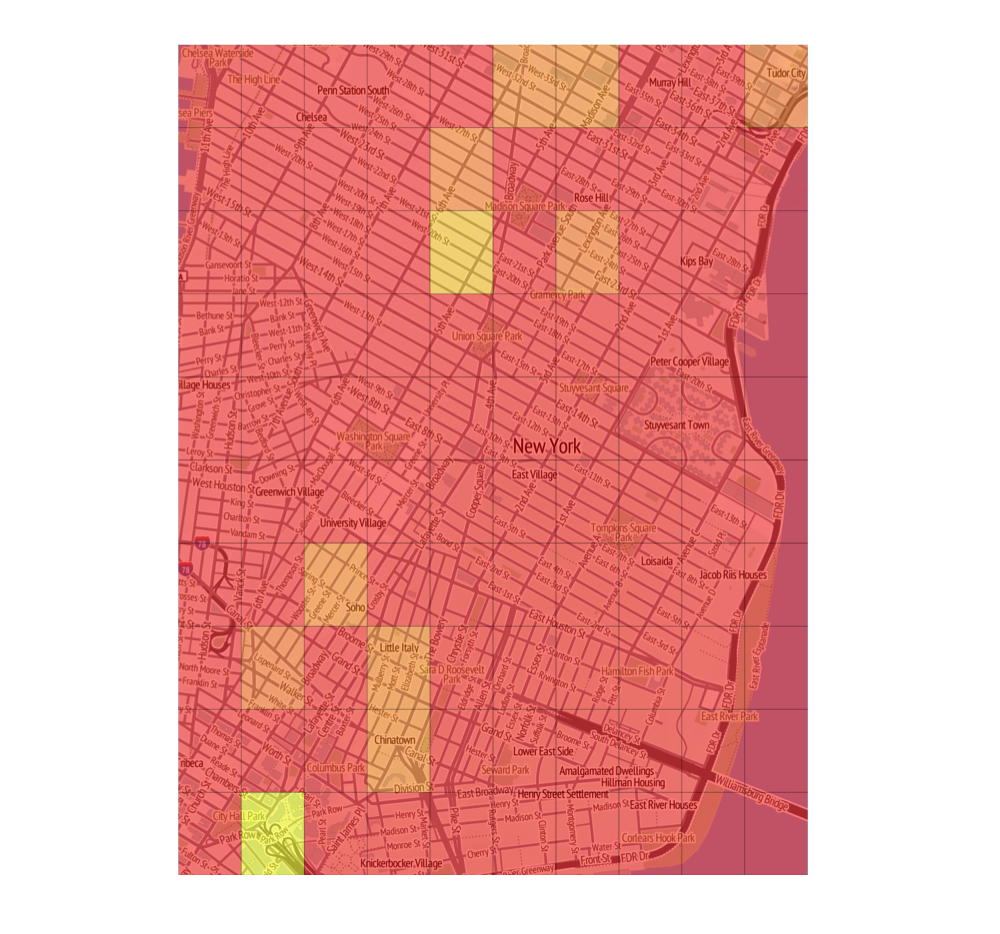}}
  \subfloat[``Went to a great Italian restaurant"]{\label{fig:rest3}\includegraphics[scale=0.28]{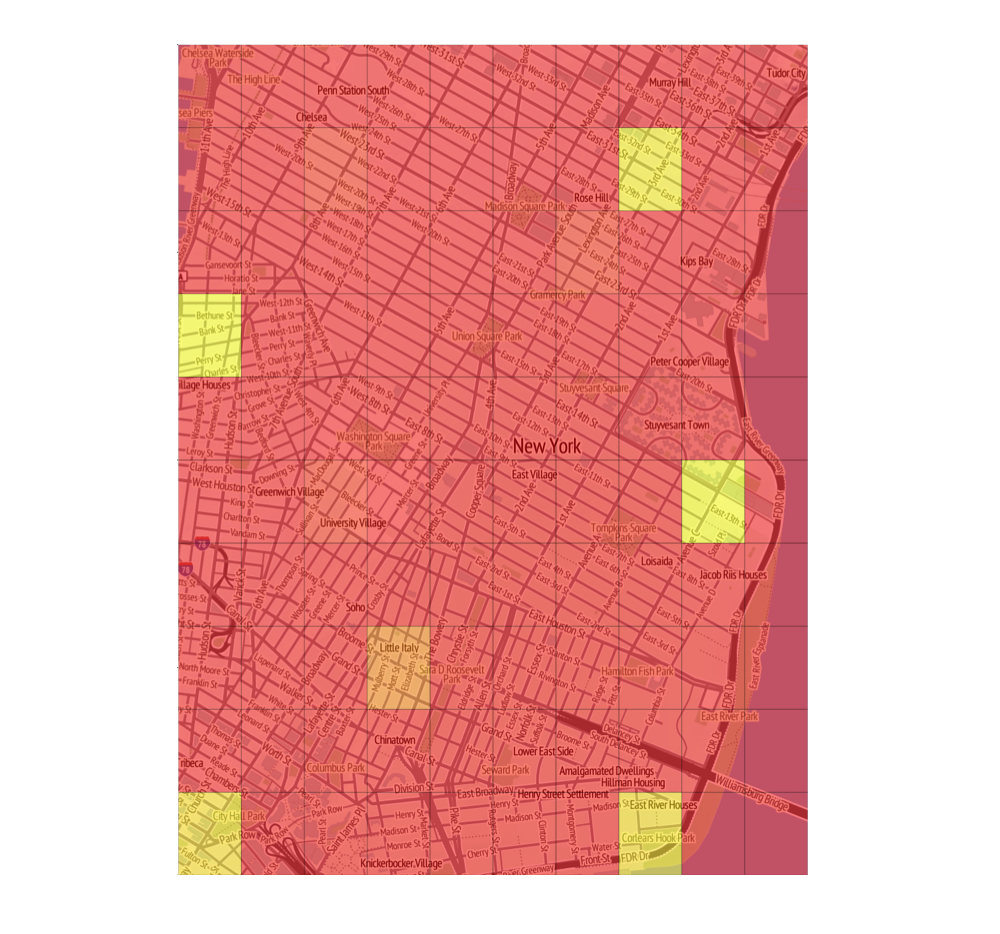}}
  \caption{Local-level Perceptions in NYC}
  \label{fig:rest}
\end{figure*}

We harvested geo-tagged tweets from Twitter
between January $29^{th}$ and February $28^{th}$ 2013 
from three areas, namely, Downtown Manhattan in New 
York City (NYC), the greater Washington D.C. area and 
its surroundings (DC), and the entire state of Connecticut (CT). 
Although these areas can be partitioned according to town and
city jurisdictions or even by zip code, for the sake
of illustration, we divide them into $100$ equal sub-areas along a 
$10 \times 10$ grid using latitudinal and longitudinal coordinates. 
The partitions of each area differ widely: (i)~NYC sub-areas 
include a few blocks and provide a {\em local-level} perspective;
(ii)~DC sub-areas include substantial portions of cities, suburbs, 
and interstates providing a {\em district-level} perspective;
and (iii)~CT sub-areas contain multiple towns,
entire cities and woods offering a {\em region-level} perspective. 

For each area, we eliminated non-English tweets and those without 
geo-tags. Table~\ref{tab:summary} shows that
the tweet density in NYC is an order higher than DC and two orders higher 
than CT. The lower densities in DC and CT, however, do not impede 
training of the language models, because in each area 
tweet distributions conform to population spread as shown in 
Figure~\ref{fig:locs}~\cite{density}. Thus, tweets in NYC are almost 
uniform, in DC they cluster around major cities and follow paths to 
major highways, and in CT they concentrate around the 
three major interstates, with sparse densities in the woods and 
farmland towns. 

\begin{figure*}[!ht]
\vspace*{-0.1in}
  \centering
  \subfloat[``traffic'']
  {\label{fig:traf1}\includegraphics[scale=0.3]{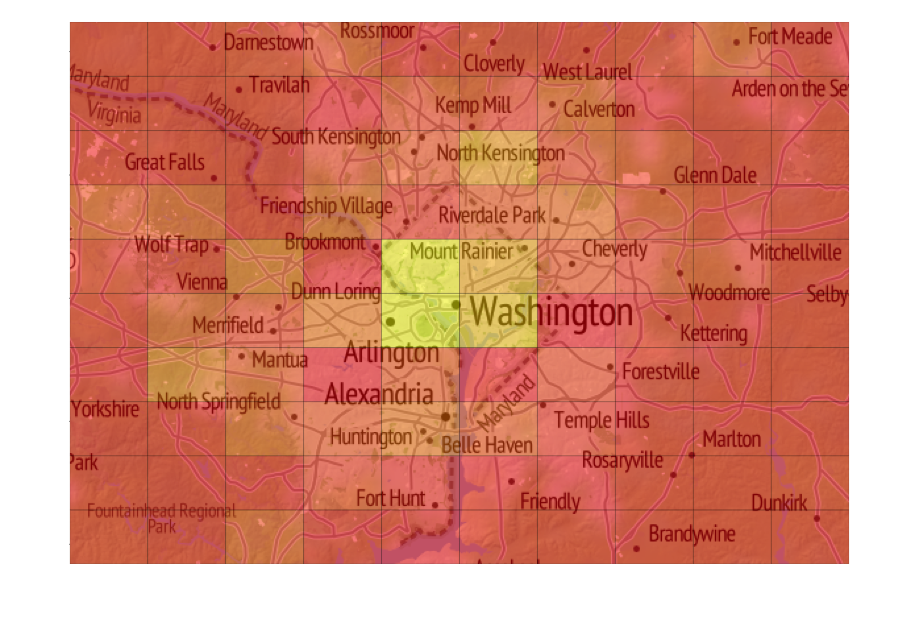}}
  \subfloat[``traffic during commute'']
  {\label{fig:traf2}\includegraphics[scale=0.3]{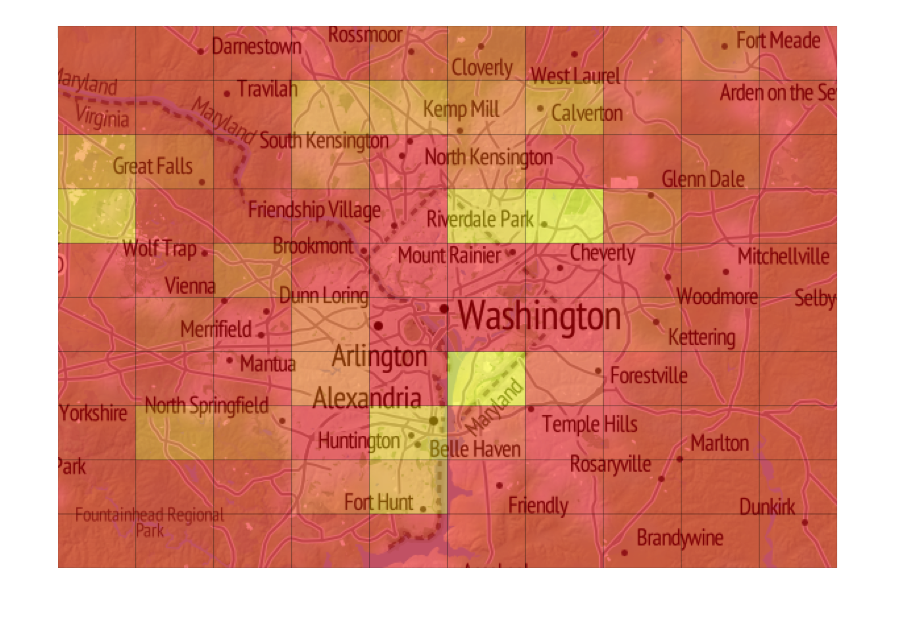}}
  \caption{District-level Perceptions in DC}
  \label{fig:traf}
\vspace*{-0.1in}
\end{figure*}

\begin{table}[!ht]
\vspace*{-0.03in}
\begin{center}
\begin{tabular}{|c|c|c|c|c|} \hline
Area & Sub-area & Area Size & Tweets & Density \\ \hline
 NYC & Local & 82.3km$^2$ & 110,924 & 1,347/km$^2$ \\
 DC & District & 3,452km$^2$ & 394,072 & 114/km$^2$ \\
 CT & County & 22,140km$^2$  & 355,678 & 16/km$^2$ \\ \hline
\end{tabular}
\end{center}
\vspace*{-0.2in}
\caption{Area-wise Summary of Tweets}
\label{tab:summary}
\vspace*{-0.03in}
\end{table}

Geo-tagged tweets were further pre-processed by converting all words to 
lowercase and by stripping punctuation, hashtags (terms starting with \#), 
username replies (terms beginning with $@$), and Web links. 
Common words such as ``at", ``the'', and ``or'' lack contextual information,
and hence, were eliminated using a stopword list of $200$ most 
frequently used words. The stopword list was limited to $200$ 
which is approximately equal to $1\%$ of the average number 
of distinct words across each area.  We also include a ``catch all'' 
unigram ``$<$misc$>$" to aggregate the probability of all words that occur 
only once. It also accounts for miscellaneous, shorthand,
mis-spelled, and other user-specific notations.
On an average $3.16\%$ of the words in each area were mapped to ``$<$misc$>$'',
suggesting that we can control this source of distortion 
without impacting the models' fidelity. 

\section{Illustrations}
\label{sec:ill}
In this section, we illustrate how our approach can identify sub-areas 
with strong topic-based perceptions.

\subsection{Perceptions in NYC}
Downtown Manhattan is a popular tourist destination and includes 
Chinatown and Little Italy as well as Broadway and Penn Station. 
Given its multi-cultural 
neighborhoods and popularity, this area is rife with many types of eateries,
due to which we extract perceptions about restaurants. 
Figure~\ref{fig:rest1} displays the results in the form of a 
heat map, produced by a generic query ``restaurants''. 
Brighter shades across many sub-areas indicate that people discuss 
restaurants broadly. Figure~\ref{fig:rest2} shows the results of a 
refined query ``Italian restaurants''. The heat map now 
concentrates on fewer sub-areas, mostly in the southwest,
which corresponds to Little Italy. It also includes 
northern sub-areas; home to many high-end Italian 
restaurants\footnote{\url{http://www.zagat.com}}. Finally, a 
specific query ``went to a great Italian restaurant'' produces 
Figure~\ref{fig:rest3} which tells us that this perception 
is most strongly present in Little Italy, and at the entrances to 
the Holland Tunnel, Brooklyn and Williamsburg Bridges. That this perception
is strong in sub-areas used to leave the city suggests 
that visitors may be more inclined to share their satisfaction 
about a great meal in Little Italy compared to the city's residents. 

\begin{figure*}[!ht]
\vspace*{-0.05in}
  \centering
  \subfloat[``crime'']{\label{fig:crime}\includegraphics[scale=0.3]{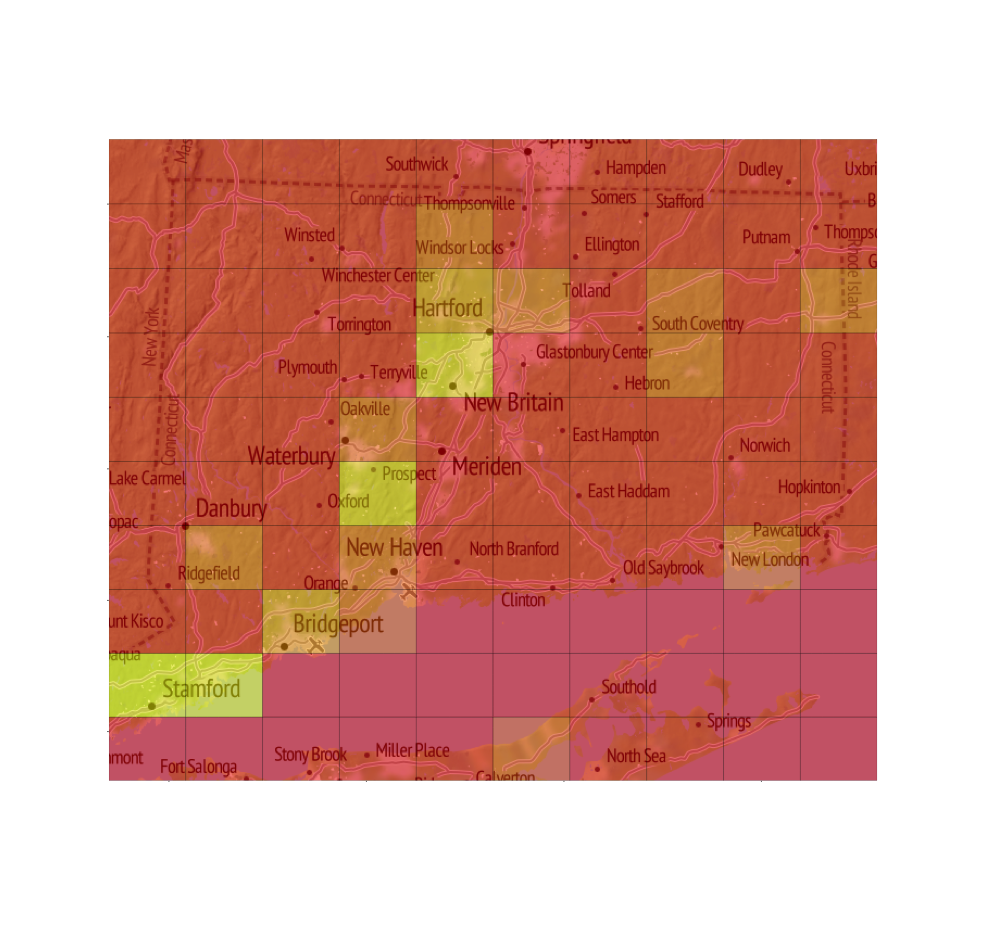}}
  \subfloat[``hospital'']{\label{fig:hosp}\includegraphics[scale=0.3]{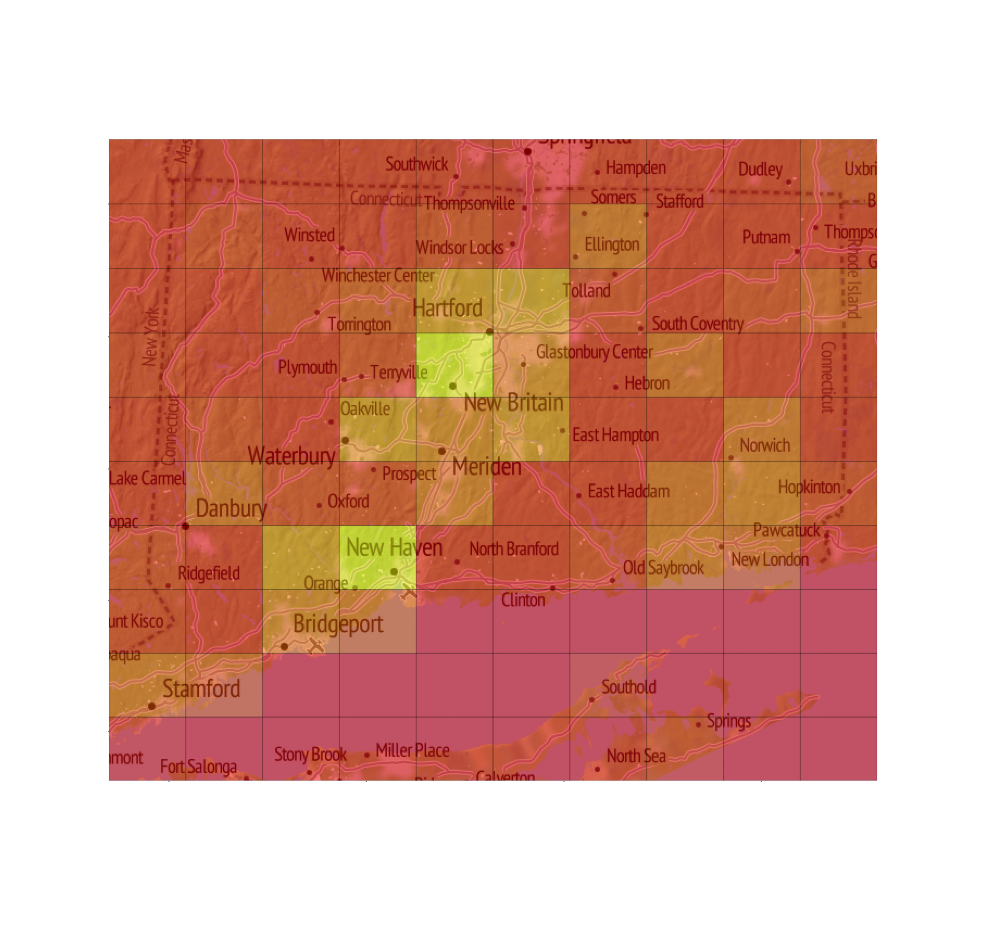}}
  \caption{Region-level Perceptions in CT}
  \label{fig:ct1}
\vspace{-0.05in}
\end{figure*}

\subsection{Perceptions in DC}
This area encompasses Washington D.C. and the surrounding 
suburbs. Here, sub-areas include
entire communities, parts of Washington D.C., portions of the
I-95/495 interstate loop infamous for its heavy traffic, 
regional parks, and major
roads that connect Washington D.C. to Maryland and Virginia. The city
is dominated by office parks, federal agencies, and corporate headquarters
bringing in a large number of commuters from outside towns and suburbs. 
We thus extract perceptions on ``traffic'' for this area.
The heat map in Figure~\ref{fig:traf1}, resulting from a generic 
query ``traffic'', shows that traffic is most strongly perceived 
inside and around the four sub-areas of downtown and decreases 
in prominence as we go farther away. The heat map in Figure~\ref{fig:traf2}, 
resulting from a more nuanced query ``traffic during commute'', 
finds that people do not discuss traffic and commute within
the city, but as expected in the sub-areas which contain portions of 
the I-95/495 interstate loop and those to the west neighboring 
Dulles Airport. 

\subsection{Perceptions in CT}
This area covers the state of Connecticut featuring 
large sub-areas that include entire towns and cities. 
A hot-button issue that many people consider when deciding to
relocate to a neighborhood is the public perception of crime. 
City and town governments must thus be aware of how crime 
is perceived in their jurisdictions. The heat map in Figure~\ref{fig:crime}, 
resulting from a generic query ``crime'', suggests that the people of 
CT do not think about crime except for sub-areas along the I-91 interstate 
containing the cities of New Haven, Bridgeport, Stamford, and Hartford,
which are notorious for its dangerousness\footnote{\url{http://www.fbi.gov/about-us/cjis/ucr/crime-in-the-u.s/2012}}. Because a state 
encompasses a large area, 
we also extract perceptions about topics that are less 
likely to be thought of at a local- and district-level. The heap map in 
Figure~\ref{fig:hosp}, 
resulting from the query ``hospital'', shows that the hottest sub-areas 
coincide with the Yale-New Haven Hospital and UConn Health Center.  
Also, adjacent sub-areas are more likely to 
think of hospitals, compared to other sub-areas in the state. 

\section{Storm Power Grid Damage Response}
\label{sec:appl}
\begin{figure}
\includegraphics[scale=0.5]{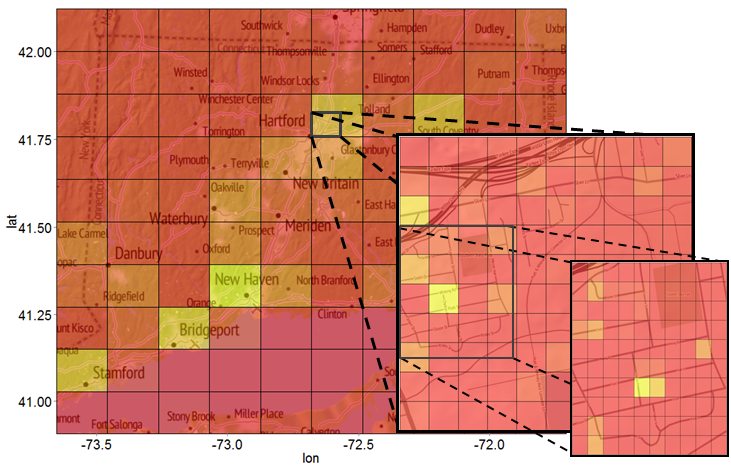}
\caption{Perception of ``power outage'' in CT}
\label{fig:storm}
\end{figure}

The above examples illustrate the capability of our 
approach to identify varied topic-based perceptions. We now
discuss how such identification can be leveraged for
a disaster response scenario. Natural and other disasters can 
create potentially life-threatening conditions 
because of their destructive impact on an electric 
power grid. Responding to such outages efficiently
and quickly can minimize this damage and reduce the 
costs of restoration and loss of productivity. Currently, utilities 
rely on experts to estimate locations of outages and 
the type and extent of damages in order to dispatch 
appropriate crews and materials.  
We describe how public perceptions on the damages caused 
to the power grid following a storm provides 
situational or ``on-the-ground" data supporting the expert's analysis.
On January 31$^{st}$ 2013, the CT area experienced hurricane 
force winds causing widespread power outages. 
The heat map in Figure~\ref{fig:storm}, resulting from
the query ``power outage'', shows that locations of this
perception correspond to the 
power outage map~\footnote{\url{http://www.ctpost.com/local/article/Storm-leaves-thousands-without- power-4238526.php}}. Zooming into an area 
northeast of Hartford, generates another heat map for the same perception,
which now highlights a residential block. Zooming in even further 
identifies specific places on the streets, which may correspond to houses 
or electric poles. Based on this data, experts can tag
these streets as prone to power loss. They can analyze the power grid
in the area to determine what components failed, and using tweets within the 
highlighted sub-areas, hypothesize about the source of damage 
(e.g. downed trees crushing overhead lines). 
The appropriate crews and components can then be dispatched to quickly repair these failures.

\section{Related Research}
\label{sec:rr}
Language models have been built over social media posts for a 
variety of purposes. Iskandar {\em et al.} develop a query likelihood model 
with Dirichlet smoothing to retrieve content from social media and 
Wikipedia articles~\cite{iskandar07}. Li predicts the point-of-interest of 
a tweet with a unigram language model~\cite{li11}. The 
models are also combined with location information to evaluate the origin of 
tweets. Kinsella {\em et al.} estimate cities from which tweets originate 
by comparing KL-divergences among language models~\cite{kinsella11}. 
Chandra {\em et al.} also predict
originating cities based on unigram models over 
tweet-reply chains~\cite{chandra11}. Chang
predicts positions based on the spatial usage of words 
in the tweets~\cite{chang12}, 
Sadilek {\em et al.} incorporate the position of friends and content of tweets
for prediction~\cite{sadilek12}, Liu {\em et al} consider check-in histories 
with tweet content~\cite{liu12} and Wing {\em et al} use unigram models of 
tweet content across areas within geo-grids~\cite{wing11}.

This work differs from contemporary efforts because rather than focusing 
on a specific set of tasks, we develop language models to generally
identify perceptions of users across geographic
areas. Also, our sophisticated language model uses smoothing to
combine accurate estimation of unigrams with 
contextual information in bigrams compared to the prevalent models that 
consider only unigrams.

\section{Conclusions and Future Work}
\label{sec:conc}
This paper presented a methodology to identify where perceptions 
about a topic are strongly represented across a given geographic area. 
Central to the methodology are language models that can be queried 
using phrases that define any kind of perception for any topic. Without
any {\em a priori} information and aid of external data sources, 
we demonstrate how the approach can identify where a specific 
topic-based perception is strongly represented in sub-areas 
with sizes ranging from just a  few urban blocks to entire cities.

In the future, we plan to enhance the methodology with geographic
and temporal variations in word usage. We will also explore the 
use of the methodology for many different 
applications including location prediction, storm and disaster management, and 
analytics for city planning and public services including mass transit.

\bibliographystyle{abbrv}
\bibliography{seke13}
\end{document}